\documentclass[usegraphicx,usenatbib]{mn2e}
\usepackage{amsmath}
\usepackage{times}
\usepackage{subfigure}
\usepackage{epsfig}

\title[A New Algorithm For Difference Image Analysis]
  {A New Algorithm For Difference Image Analysis}

\author[D.M. Bramich]
  {D.M.~Bramich$^{1}$\thanks{E-mail: dmb@ing.iac.es}
 \medskip  
 \\$^1$Isaac Newton Group of Telescopes, Apartado de Correos 321,
       E-38700 Santa Cruz de la Palma, Canary Islands, Spain}

\begin{document}

\date{Accepted . Received ; in original form }

\pagerange{\pageref{firstpage}--\pageref{lastpage}} \pubyear{2004}

\maketitle

\label{firstpage}

\begin{abstract} 

In the context of difference image analysis (DIA), 
we present a new method for determining the convolution kernel matching a pair of images of the same field. 
Unlike the standard DIA technique which involves 
modelling the kernel as a linear combination of basis functions, we consider the kernel as a discrete 
pixel array and solve for the kernel pixel values directly using linear least-squares.
The removal of basis functions from the kernel model is advantageous for a number
of compelling reasons. Firstly, it removes the need for the user to specify such functions,
which makes for a much simpler user application and avoids the risk of an inappropriate choice.
Secondly, basis functions are constructed around the origin of the kernel coordinate system, 
which requires that the two images are perfectly aligned for an optimal result. The pixel kernel model is 
sufficiently flexible to correct for image misalignments, and in the case of a simple translation between images,
image resampling becomes unnecessary. Our new algorithm can be extended to spatially varying kernels by
solving for individual pixel kernels in a grid of image sub-regions and interpolating the solutions to obtain the kernel
at any one pixel.

\end{abstract} 


\section{Introduction}

Difference image analysis (DIA) has rapidly moved to the forefront of modern
techniques for making time-series photometric measurements on digital images.
The method attempts to match one image to another by deriving a convolution
kernel describing the changes in the point spread function (PSF) between images.
When applied to a time-series of images using a high signal-to-noise reference image,
the differential photometry that can be performed on the difference images
regularly provides superior accuracy to more traditional profile-fitting photometry, 
achieving errors close to the theoretical Poisson limits. Moreover, DIA is the only reliable 
way to analyse the most crowded stellar fields.

One will find DIA in use in many projects studying object variability. For example, microlensing searches 
(e.g. \citealt{bon2001}; \citealt{woz2001}) have been revolutionised by the ability of DIA to 
deal with exceptionally crowded fields, and surveys for transiting planets
(e.g. \citealt{bra2005}; \citealt{moc2005}) looking for small $\sim$1\% photometric eclipses 
have benefited substantially from the extra accuracy obtained with this method. Also, DIA is not limited to 
stellar photometry as illustrated by the discovery of light echoes from three ancient supernovae
in the Large Magellanic Cloud (\citealt{res2005}).

The first attempts at image subtraction are summarised in the introduction of \citet{ala1998} (from now on AL98) and are based
on trying to determine the convolution kernel by taking the ratio of the Fourier transforms of matching bright
isolated stars on each image (\citealt{tom1996}). Development of DIA reached an important landmark in AL98 with their 
algorithm to determine the convolution kernel directly in image space (rather than Fourier space) 
from all pixels in the images by decomposing the kernel onto a set of basis functions. The algorithm is
very successful and efficient, and with the extension to a space-varying kernel solution described in \citet{ala2000}
(from now on AL00), the method has become the current standard in DIA. In fact, all DIA packages use the associated 
software package ISIS2.2\footnote{http://www2.iap.fr/users/alard/package.html} (e.g. \citealt{woz2000}; \citealt{gos2002}), 
or are implementations of the Alard algorithm (e.g. \citealt{bon2001}). We refer to the method described in
AL98 and AL00 as the Alard algorithm.

In this letter we suggest a change to the main algorithm to determine the convolution kernel that
retains the linearity of the least-squares problem and yet is simpler to implement, has fewer
input parameters and is in general more robust (Section~2). We compare our algorithm
directly to the Alard algorithm (Section~3), and suggest more techniques that increase the quality of the subtracted images.
We conclude in Section~4.

\section{A New Approach To The Kernel Solution}

\subsection{Motivation}

Consider a pair of registered images of the same dimensions, one being the reference image with pixels $R_{ij}$, and the other 
the current image to be analysed with pixels $I_{ij}$, where $i$ and $j$ are pixel indices
refering to the column $i$ and row $j$ of the image. Ideally the reference image will be the better 
seeing image of the two and have a very high signal-to-noise ratio. This can be achieved in practice by stacking
a set of best-seeing images.

As with the method of AL98, we use the model
\begin{equation}
M_{ij} = [R \otimes K]_{ij} + B_{ij}
\label{eqn:model}
\end{equation}
to represent the current image $I_{ij}$, where we wish to find a suitable convolution kernel $K$
and differential background $B_{ij}$.
Formulating this as a least-squares problem, we want to minimise the chi-squared
\begin{equation}
\chi^{2} = \sum_{i,j} \left( \frac{ I_{ij} - M_{ij} }{ \sigma_{ij} } \right)^{2}
\label{eqn:chisq}
\end{equation}
where the $\sigma_{ij}$ represent the pixel uncertainties.

At this point in the Alard algorithm, the problem is converted to standard linear least-squares
by decomposing the kernel $K$ onto a set of gaussian basis functions, each multiplied by
polynomials of the kernel coordinates $u$ and $v$, and by assuming that the differential background $B_{ij}$
is represented by a polynomial function of the image coordinates $x$ and $y$.
Spatial variation of the convolution kernel is modelled by further multiplying the kernel basis functions by 
polynomials in $x$ and $y$.

This method has a major drawback in that it assumes that the chosen kernel decomposition is sufficiently complex so as
to model in detail the convolution kernel. How do we know that we are making the correct choice of
basis functions? Different situations may require different combinations of
basis functions of varying complexity. In fact, a feature of all current DIA packages (which are all based on 
the AL98 prescription for kernel basis functions) is the requirement that the
user defines the number of gaussian basis functions used, their associated sigma values and
the degrees of the modifying polynomials. This sort of parameterisation can end up being confusing for  
the user, and require a large amount of experimentation to obtain the optimal result for a specific data
set.

\subsection{Solving For A Spatially Invariant Kernel Solution}

With this motivation, we have developed a new DIA algorithm in which we make no assumptions about
the functional form of the basis functions representing the kernel.
Considering a spatially invariant kernel, we represent the kernel 
as a pixel array $K_{lm}$ with $N_{K}$ pixels
where $l$ and $m$ are pixel indices corresponding to the column $l$ and row $m$ of the kernel.
We also define the differential background as some unknown constant $B_{0}$. 
Hence we may rewrite equation~(\ref{eqn:model}) as:
\begin{equation}
M_{ij} = \sum_{l,m} K_{lm} R_{(i+l) (j+m)} + B_{0}
\label{eqn:mod_const}
\end{equation}
This equation has $N_{K} + 1$ unkowns for which we require a solution.
Note that the kernel may be of any shape that includes the pixel $K_{00}$, and so to preserve symmetry in all directions,
we adopt a circular kernel (instead of the standard square shape).

In order to solve for $K_{lm}$ and $B_{0}$ in the least-squares sense, we note that the $\chi^{2}$ in equation~(\ref{eqn:chisq})
is at a minimum when the gradient of $\chi^{2}$ with respect to each of the parameters $K_{lm}$ and $B_{0}$ is equal to zero. 
Performing the $N_{K} + 1$ differentiations and rewriting the set of linear equations in matrix form, we obtain 
the matrix equation $\mathbf{U} \mathbf{a} = \mathbf{b}$ with:
\begin{multline}
U_{pq} = 
\begin{cases}
\sum_{i,j} \frac{ R_{(i+l) (j+m)} R_{(i+l^{\prime}) (j+m^{\prime})} }{\sigma_{ij}^2}    
           & \text{for $1 \leq p \leq N_{K}$ and $1 \leq q \leq N_{K}$} \\
\sum_{i,j} \frac{ R_{(i+l^{\prime}) (j+m^{\prime})} }{\sigma_{ij}^2}    
           & \text{for $p = N_{K} + 1$ and $1 \leq q \leq N_{K}$} \\
\sum_{i,j} \frac{ R_{(i+l) (j+m)} }{\sigma_{ij}^2}
           & \text{for $1 \leq p \leq N_{K}$ and $q = N_{K} + 1$} \\
\sum_{i,j} \frac{1}{\sigma_{ij}^2}     
           & \text{for $p = q = N_{K} + 1$} \\ 
\end{cases} \\
a_{p} = 
\begin{cases} 
K_{lm} & \text{for $1 \leq p \leq N_{K}$} \\
B_{0} & \text{for $p = N_{K} + 1$} \\
\end{cases} \\
b_{p} = 
\begin{cases}
\sum_{i,j} \frac{ I_{ij} R_{(i+l) (j+m)} }{\sigma_{ij}^2}
           & \text{for $1 \leq p \leq N_{K}$} \\
\sum_{i,j} \frac{ I_{ij} }{\sigma_{ij}^2}
           & \text{for $p = N_{K} + 1$} \\
\end{cases} \\
\label{eqn:matrix}
\end{multline}
where $p$ and $q$ are generalised indices for the vector of unknown quantities
$\mathbf{a}$, with associated kernel indices $(l,m)$ and $(l^{\prime},m^{\prime})$
respectively. Finding the solutions\footnote{Iteration is required for a 
self-consistent solution for $K_{lm}$ and $B_{0}$ since the solution
depends on the pixel variances $\sigma_{ij}^{2}$ which in turn depend
on the image model values $M_{ij}$. See Section~\ref{uncertainties}}
for $K_{lm}$ and $B_{0}$ requires the construction of the matrix
$\mathbf{U}$ and vector $\mathbf{b}$, inverting $\mathbf{U}$ and calculating 
$\mathbf{a} = \mathbf{U^{-1}} \mathbf{b}$. 

Every pixel on both the reference image and current image has the potential to be included in
the calculation of $\mathbf{U}$ and $\mathbf{b}$. However, we ignore bad/saturated pixels on 
both images, and also any pixels on the current image for which the calculation of the 
corresponding model pixel value includes a bad/saturated pixel on the reference image.
This implies that a single bad/saturated pixel on the reference image can discount a set of pixels
equal to the kernel area on the current image. Hence bad/saturated pixels on the reference image
should be kept to a minimum, and excessively large kernels should be avoided.

The kernel sum $P = \sum_{l,m} K_{lm}$ is a measure of the mean scale factor between the
reference image and the current image, and consequently it includes the effects of 
relative exposure time and atmospheric extinction. We refer to $P$ as the photometric scale factor. Although it is not
essential, we suggest that a constant background estimate is subtracted from the reference image before
solving for the kernel and differential background since this will
minimse any correlation between $P$ and $B_{0}$.

Finally, we mention that a difference image $D_{ij}$ is defined as $D_{ij} = I_{ij} - M_{ij}$. 
Assuming that most objects in the reference image are constant sources, then a difference image
will consist of random noise (mainly Poisson noise from photon counting) except where a source has varied in brightness
or the background pattern has varied. Sources that are brighter or dimmer at the epoch of the current image relative
to the epoch of the reference image will show up as positive or negative flux residuals, respectively, on the difference image.
These areas may be measured to yield a difference flux for each object of interest.

\subsection{Uncertainties Arise From The Model, Not The Data \label{uncertainties}}

We take the following standard CCD noise model for the pixel variances:
\begin{equation}
\sigma_{ij}^{2} = \frac{\sigma_{0}^{2}}{F_{ij}^{2}} + \frac{M_{ij}}{G F_{ij}}
\label{eqn:noise_model}
\end{equation}
where $\sigma_{0}^{2}$ is the CCD readout noise (ADU), $G$ is the CCD gain (e$^{-}$/ADU) and 
$F_{ij}$ is the master flat field image. Note that the $\sigma_{ij}$ depend on the image model $M_{ij}$
and consequently, fitting $M_{ij}$ becomes an iterative process.
Note also that we assume that the reference image $R_{ij}$ 
and master flat field image $F_{ij}$ are noiseless since these are high S/N ratio images. Finally, if
the current image was registered with the reference image via a geometric transformation, then 
the flat field $F_{ij}$ that is actually used in the noise model must be the result of the same 
transformation applied to the original master flat field.

In order to calculate an initial kernel and differential background solution, we set the $M_{ij}$ to the
image values $I_{ij}$. In subsequent iterations, we use the current image model 
to set the $\sigma_{ij}$ as per equation~\ref{eqn:noise_model}. We also employ a 3$\sigma$ clip 
algorithm during the iterative model fitting process in order to prevent outlier 
image pixel values from entering the solution. After each iteration, we calculate the absolute normalised residuals
$r_{ij} = \left| D_{ij} / \sigma_{ij} \right|$ for all pixels. Any pixels with 
$r_{ij} \geq 3$ are ignored in subsequent iterations. The iterations are stopped when
no more image pixels are rejected and at least two iterations have been performed.

\begin{figure*}
\def\subfigtopskip{4pt}
\def\subfigbottomskip{8pt}
\def\subfigcapskip{4pt}
\centering
\epsfig{file=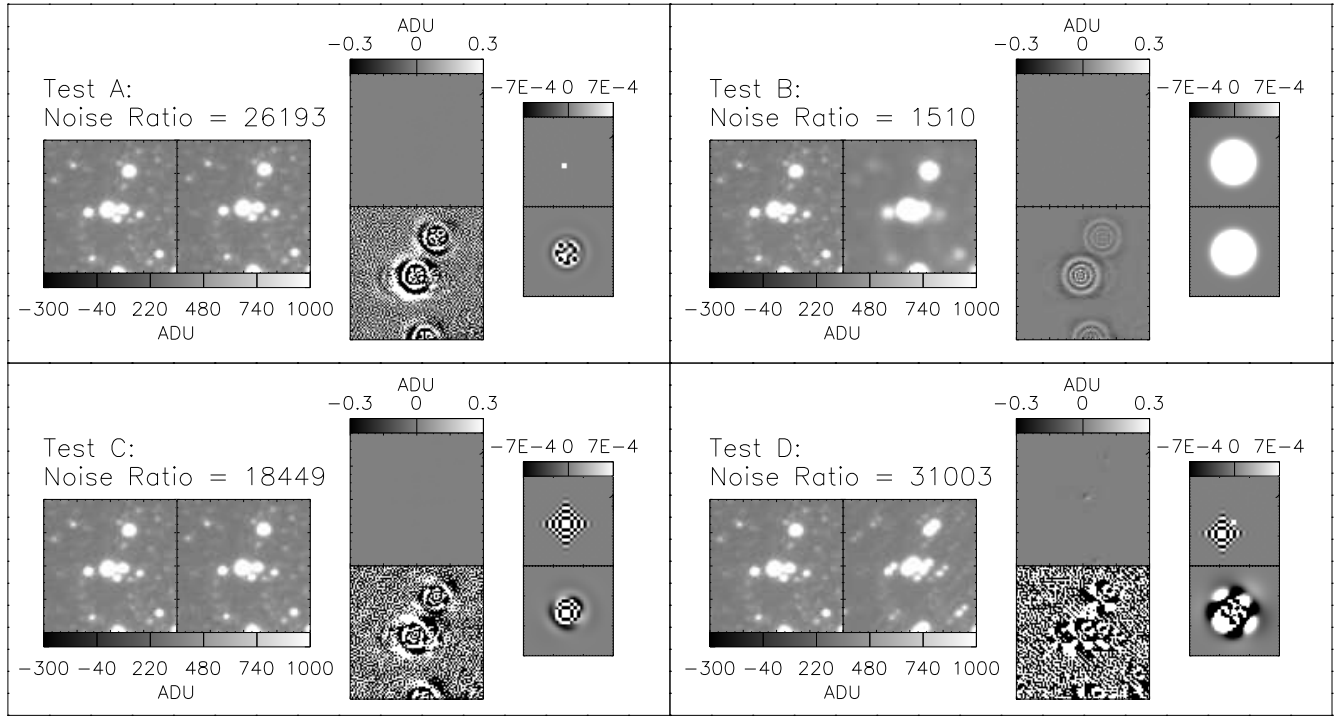,width=\linewidth}
\caption{Shown are four difference imaging tests A-D. In each panel corresponding to a test,
         there are three pairs of images, where each pair is shown using the same intensity
         scale indicated by the graduated colour bar. In the left most pair, the reference 
         image and current image sub-regions are shown on the left and right, respectively.
         In the middle pair, the DANDIA and ISIS2.2 difference images are shown
         at the top and bottom, repsectively. In the right most pair, the corresponding
         DANDIA and ISIS2.2 kernel solutions are shown at the top and bottom, respectively.
         \label{fig:test1}}
\end{figure*}

\subsection{Solving For A Spatially Variant Kernel Solution}

In extending our new method to solving for a spatially variant kernel solution, we preserve
flexibility by splitting the image area into an $N_{x} \geq 1$ by $N_{y} \geq 1$ grid of sub-regions
and solving for the kernel and differential background in each sub-region. The coarse grid of kernel
and differential background solutions may be interpolated to yield the solution corresponding to
any given image pixel. In this way we make no assumptions about how
the kernel and differential background vary across the image area. This is in contrast to AL00, whose method  
employs an extension of the kernel basis functions by further multiplication by polynomials in $x$ and $y$, and
therefore requires two more input parameters from the user, namely the degrees of the polynomials describing the
spatial variation of the kernel and the differential background.

\section{Comparisons With The Alard Algorithm}

\subsection{Initial Tests \label{initialtests}}

To illustrate the potential advantages of our new kernel solution method over that of AL98, we carry out a set of
simple tests on a 1024$\times$1024 pixel CCD image of the globular cluster NGC1904.
In each test we use the original image as the reference image $R_{ij}$ and a transformed version of the original 
image as the current image $I_{ij}$, where the transformations employed are simple, spatially invariant and typical of astronomical 
imaging. We attempt to solve for the kernel using our new method, which is implemented in a
software package called DANDIA (Bramich in prep.), and we compare the solution to that obtained
using the ISIS2.2 software from AL00. We use the ISIS2.2 default parameters
specifying 3 gaussian basis functions of $\sigma = 0.7, 2.0, 4.0$~pix with modifying polynomials of degree
6, 4 and 3, respectively. For both software packages, we choose to solve for a spatially invariant kernel 
of size 27$\times$27 pixels, and a constant differential background. 

The better the match between the convolved reference image and the current image, the smaller the value of the quantity
$S^{2} = \sum_{i,j} D_{ij}^{2}$. We guage the relative quality of the kernel 
solutions by calculating the noise ratio $S_{\mbox{\footnotesize ISIS}} / S_{\mbox{\footnotesize DANDIA}}$ where 
$S_{\mbox{\footnotesize ISIS}}$ and $S_{\mbox{\footnotesize DANDIA}}$ are values of $S$
calculated for a small 80x80 pixel sub-region using ISIS2.2 and DANDIA, respectively. 

The results of the tests described below are shown in Figure~\ref{fig:test1}:
\begin{enumerate}
\item In test A, the current image has been created by shifting the reference image by one pixel
      in each of the positive $x$ and $y$ spatial directions, without resampling. 
      The corresponding kernel should be the identity kernel (central pixel value of 1 and 0 elsewhere)
      shifted by one pixel in each of the negative $u$ and $v$ kernel coordinates. DANDIA recovers this kernel
      to within numerical rounding errors whereas ISIS2.2 recovers a peak pixel value of 0.995 with other absolute pixel 
      values of up to 0.004. Consequently the residuals in the ISIS2.2 difference image are considerably worse than
      those for DANDIA, and the noise ratio
      is $S_{\mbox{\footnotesize ISIS}} / S_{\mbox{\footnotesize DANDIA}} \approx 26190$.
\item In test B, the current image has been created by convolving the reference image with a gaussian
      of FWHM 4.0~pix. Both DANDIA and ISIS2.2 recover the kernel successfully, but DANDIA 
      out-performs ISIS2.2 with $S_{\mbox{\footnotesize ISIS}} / S_{\mbox{\footnotesize DANDIA}}~\approx~1510$.
\item In test C, we shifted the reference image by half a pixel in each of the positive $x$ and $y$
      spatial directions to create the current image, an operation that required the resampling of the reference
      image. We used the cubic O-MOMS resampling method (see Section~\ref{resample}). ISIS2.2 fails to reproduce the highly
      complicated kernel matching the two images, whereas DANDIA does a nearly perfect job. The noise ratio is
      $S_{\mbox{\footnotesize ISIS}} / S_{\mbox{\footnotesize DANDIA}} \approx 18450$.
\item In test D, we simulate a telescope jump by setting 
      $I_{ij}~=~\left(0.6~\times~R_{ij}\right)~+~\left(0.4~\times~R^{\prime}_{ij}\right)$
      where $R^{\prime}_{ij}$ is a resampled version of the reference image shifted by 3.5 pixels in each of the
      positive $x$ and $y$ spatial directions. The corresponding kernel is a combination of the identity kernel
      and a shifted version of the kernel from test~C. DANDIA accurately reproduces this kernel with a 
      central pixel value of 0.60015 whereas ISIS2.2 produces a poor approximation of the kernel with a central 
      pixel value of 0.631. The noise ratio is $S_{\mbox{\footnotesize ISIS}} / S_{\mbox{\footnotesize DANDIA}} \approx 31000$.
\end{enumerate}

It is evident that the gaussian basis functions used in ISIS2.2 limit the flexibility of the kernel solution to 
modelling kernels that are centred near the kernel centre and that have scale sizes similar to the sigmas of the gaussians
employed. It is only in test B that ISIS2.2 can closely model the kernel, simply because the kernel itself is
a gaussian. Tests A, C \& D show how ISIS2.2 is unable to model sharp, complicated and off-centred kernels.
DANDIA does not suffer from any of these limitations since it makes no assumption about the kernel shape,
and hence it performs superbly in all of the above tests.

\subsection{Image Resampling \label{resample}} 
      
In Section~2, we make the assumption that the reference image and current image are registered, which
implies that one of the images has been transformed to the pixel coordinate system of the other image, usually 
via image resampling. Ideally one should
transform the reference image to the current image since the reference image forms part of the model. In this way, the pixel
variances in the current image are left uncorrelated from pixel to pixel. However, most implementations of 
DIA transform the current image to the coordinate system of the reference image using image resampling. 

We suggest two improvements to this methodology. Firstly, if resampling is to be employed, one should use an optimal resampling method.
We employ the cubic O-MOMS (Optimal Maximal-Order-Minimal-Support) basis function for resampling, which is 
constructed from a linear combination of the cubic B-spline function and its derivatives. 
The O-MOMS class of functions have the highest approximation order and smallest approximation error constant for a given support
(\citealt{blu2001}).

Secondly, our kernel model does not use basis functions that are functions of the kernel pixel
coordinates. Consequently, for two images that require only a translation to be registered, the image
resampling is incorporated in the kernel solution, avoiding the problem of correlated pixel noise.
DIA is used extensively for extracting lightcurves of objects in time-series images, which usually
only have a small pixel shift between images. By translating the current image to the reference
image by an integer pixel shift, avoiding image resampling, the kernel solution process
can do the rest of the job of matching the reference image to the current image.

\subsection{Final Tests}

\begin{figure*}
\def\subfigtopskip{4pt}
\def\subfigbottomskip{8pt}
\def\subfigcapskip{4pt}
\centering
\epsfig{file=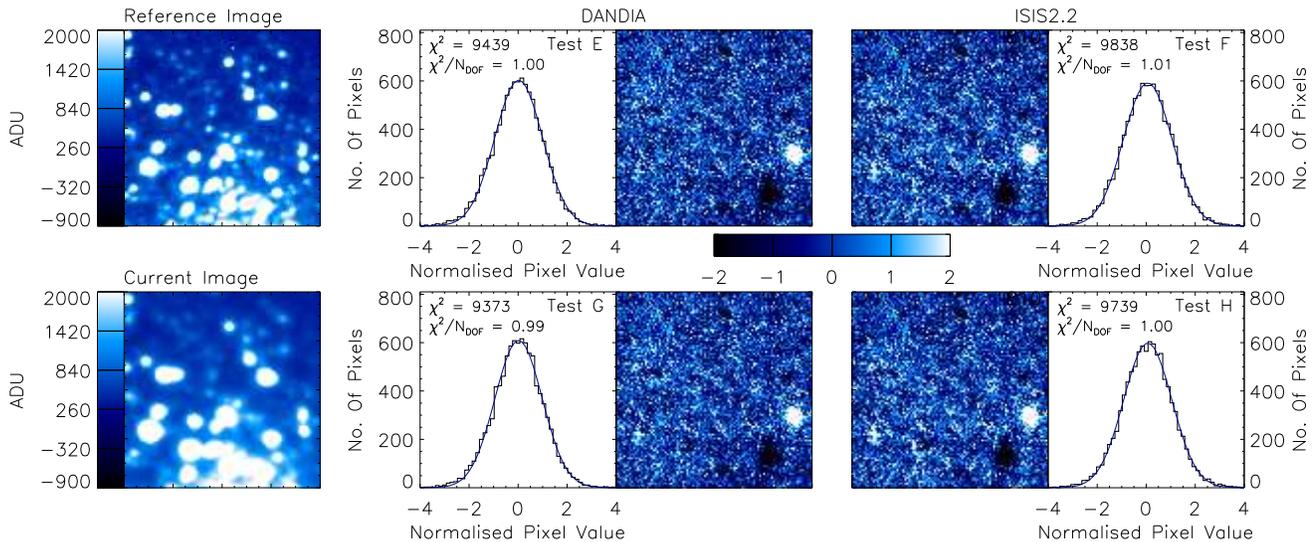,width=\linewidth}
\caption{Shown are four more difference imaging tests E-F. On the left we show 100$\times$100 pixel cutouts from the reference image
         and current image. The remaining panels show corresponding difference images normalised by the pixel noise model
         (equation~\ref{eqn:noise_model2}) with a linear scale from -2 to 2, and histograms of the normalised pixel values overlaid
         with a gaussian fit.
         \label{fig:test2}}
\end{figure*}

We now test our new algorithm on a pair of 1024$\times$1024~pixel images of NGC1904 from the same camera 
with FWHMs of $\sim$3.2~pix and $\sim$4.9~pix.
Using matching star pairs, we derive a linear transformation between the images
that consists of a translation with negligible rotation, shear and scaling.
From the calibration images, we measure a gain of 1.48~e$^{-}$/ADU and a readout noise of 4.64~ADU, and we construct
a master flat field for use in the noise model. 
On the left of Figure~\ref{fig:test2}, we present 100$\times$100 pixel cutouts of the reference image (the better seeing image) 
and the current image.

When calculating the $\chi^{2}$ of the difference images,
we use a modified version of equation~\ref{eqn:noise_model} to account for the noise contribution from the
single-exposure reference image:
\begin{equation}
\sigma_{ij}^{2} = \frac{\sigma_{0}^{2}}{F_{ij}^{2}} + \frac{M_{ij}}{G F_{ij}}
                + f^{2} \left( \sum_{l,m} K_{lmij}^{2} \left(\frac{\sigma_{0}^{2}}{F_{ij}^{2}} + \frac{R_{ij}}{G F_{ij}} \right) \right)
\label{eqn:noise_model2}
\end{equation}
where $K_{lmij}$ is the space variant kernel and $f$ is a factor correcting for the noise distortion from resampling
the reference image. The value of $f$ depends on the resampling method used and the coordinate transformation applied.
We calculate $f$ by generating a 1024$\times$1024 pixel image of values drawn from a normal distribution with zero mean
and unit sigma, resampling the image using the same method and transformation as that applied to the reference image,
and then fitting a gaussian to the histogram of transformed pixel values, the sigma of which indicates the value of $f$.
For cubic O-MOMS resampling and the transformation between our two test images, we obtain $f =$~0.884.

Our first pair of tests involves registering the images by resampling the reference image via cubic O-MOMS 
and then using DANDIA (test E) and ISIS2.2 (test F) to generate difference images. For DANDIA, we solve for an array of 
circular kernels corresponding to a 10$\times$10 grid of image sub-regions, where each kernel contains 317 pixels.
The kernel used to convolve each pixel on the reference image is calculated via bilinear interpolation of the array of kernels.
The results of test E are displayed in the upper middle panel of Figure~\ref{fig:test2} where we show the difference image normalised 
by the pixel noise from equation~\ref{eqn:noise_model2}
with a linear scale from -2 to 2. Two variable stars are visible (RR Lyraes) and the cosmic ray from the reference image
has created a negative flux on the difference image. In the same panel we plot the histogram of
normalised pixel values overlaid with a gaussian fit, and calculate a $\chi^{2}~\approx~9439$, ignoring the small pixel areas 
including the variable stars and the cosmic ray (250~pix). The 100$\times$100 pixel cutout
corresponds to one image sub-region used to determine a kernel solution and hence we may calculate a reduced 
chi-squared $\chi^{2} / N_{\text{DOF}}~=~1.00$ by assuming $N_{\text{DOF}}~=~9750~-~318$.

For ISIS2.2 we solve for a spatially variant kernel of degree 2 with a spatially variant differential background of degree 3 in addition
to the other default kernel basis functions (see Section~\ref{initialtests}; 328 free parameters). The results of test F are shown in 
the upper right panel of Figure~\ref{fig:test2}. We obtain $\chi^{2}~\approx~9838$, and assuming $\sim$3 free parameters per image
sub-region, we obtain $\chi^{2} / N_{\text{DOF}}~=~1.01$.

Tests G \& H involve registering the images to within 1 pixel by translating the reference image via an integer pixel shift.
Then we apply DANDIA (test G) and ISIS2.2 (test H) to obtain kernel solutions, avoiding the use of resampling.
For DANDIA we obtain $\chi^{2}~\approx~9373$, and for ISIS2.2 we obtain $\chi^{2}~\approx~9739$, with corresponding
$\chi^{2} / N_{\text{DOF}}$ of 0.99 and 1.00, respectively (see Figure~\ref{fig:test2}).

Visually, the normalised difference image cutouts in Figure~\ref{fig:test2} are very similar, and differences are only noticeable after
detailed scrutiny. However, the $\chi^{2}$ analysis reveals that our algorithm performs considerably better than the Alard algorithm
(test E performs 0.60$\sigma$ better than test F, and test G performs 0.38$\sigma$ better than test H), 
and that image resampling degrades the difference images 
(test G performs 0.48$\sigma$ better than test E, and test H performs 0.70$\sigma$ better than test F).
The highest quality difference image was produced by using DANDIA on the two images aligned to within 1 pixel but without resampling (test G,
which performs 1.08$\sigma$ better than test F).

\section{Conclusions}

We have presented a new method for determining the convolution kernel matching a best-seeing reference image to another image of the same field. 
The method involves modelling the kernel as a pixel array, avoiding the use of possibly inappropriate basis functions, 
and eliminating the need for the user to 
specify which basis functions to use via numerous parameters. For images that require a translation to be registered, the kernel pixel array
incorporates the resampling process in the kernel solution, avoiding the need to resample images, which degrades their quality and creates
correlated pixel noise. Kernels modelled by basis functions may only partly compensate for sub-pixel
translations since the basis functions are centred at the origin of the kernel
coordinates.

We have shown that our new method can produce higher quality difference images than ISIS2.2. 
Ideally the reference image should be aligned with the current image, preferably without resampling, 
but using O-MOMS resampling when necessary. The flexibility
of our kernel model allows the construction of difference images for telescope jumps, or trailed images, which is where ISIS2.2 fails.
These improvements have important implications for time-series photometric surveys. 
Better quality difference images implies more accurate lightcurves, and the increased kernel flexibility will lead to less
data loss due to telescope tracking and/or focus errors.

\section*{Acknowledgements}

D.M.~Bramich would like to thank K.~Horne and M.~Irwin for their useful advice, and A.~Arellano Ferro for 
supplying the test images. This work is dedicated to Phoebe and Chloe Bramich Mu\~niz.

\label{lastpage}

\end{document}